\documentstyle[aps,amstex,graphics,epsfig,fleqn,floats]{revtex}

\begin{document}

\title{Solving seismic wave propagation in elastic media using the
matrix exponential approach}
\author{J.S. Kole \\
\smallskip Centre for Theoretical Physics and Materials Science Centre
\\ University of Groningen, Nijenborgh 4 \\ NL-9747 AG Groningen,
The Netherlands \\ \smallskip E-mail: j.s.kole@@phys.rug.nl}
\date{dated \today}

\maketitle

\newcommand{\dd}[1]{\frac{\partial}{\partial #1}}
\def\mH{{\mathcal H}}
\def\mU{{\mathcal U}}
\def\br{{\mathbf{r}}}
\def\be{{\mathbf{e}}}
\def\bsig{{\mathbf{\sigma}}}
\def\bs{{\mathbf{s}}}
\def\bv{{\mathbf{v}}}
\def\bw{{\mathbf{w}}}
\def\fsr{\frac{1}{\sqrt{\rho}}}
\newcommand{\php}{\phantom{+}}
\newcommand{\sump}{\mathop{{\sum}'}}
\newcommand{\pht}{\phantom{T}}
\def\mO{{\mathcal O}}
\newcommand{\ee}[1]{\cdot 10^{#1}}
\def\erf{{\operatorname{erf}}}


\begin{abstract}
Three numerical algorithms are proposed to solve the
time-dependent elastodynamic equations in elastic solids. All
algorithms are based on approximating the solution of the
equations, which can be written as a matrix exponential.
By approximating the matrix exponential with a product formula, an
unconditionally stable algorithm is derived that conserves the
total elastic energy density. By expanding the matrix exponential
in Chebyshev polynomials for a specific time instance, a so-called
``one-step'' algorithm is constructed that is very accurate with
respect to the time integration. By formulating the conventional
velocity-stress finite-difference time-domain algorithm (VS-FDTD)
in matrix exponential form, the staggered-in-time nature can be
removed by a small modification, and higher order in time
algorithms can be easily derived. For two different seismic events
the accuracy of the algorithms is studied and compared with the
result obtained by using the conventional VS-FDTD algorithm.
\end{abstract}

\section{Introduction}
An important aid in the understanding of wave propagation in
inhomogeneous media is seismic forward modeling. In all but the
simplest cases, an analytical solution of the elastodynamic
equations is not available, and one must resort to numerical
solutions. For this, two main strategies can be followed: one
solves the elastodynamic equations either in the strong
formulation, where the equations of motion and boundary conditions
are written in differential form, or in the weak formulation,
where the equations of motion are given in integral form. The
latter formulation, implemented by finite element
\cite{Lys72,Bao98}, spectral element \cite{Pat84,Koma97} or finite
integral methods \cite{Fell95,Kony98}, may be preferred to deal
with complex geometries, or non-trivial free surface boundary
conditions.

The first strategy (the strong formulation) is followed in the
finite-difference approach that is based on solving either the
first-order velocity-stress differential equations
\cite{Viri84,Viri86,Leva88}, or the second order wave equation
\cite{Kell76,Dabl87,Igel95}. In the original formulation of the
velocity-stress finite-difference time-domain (VS-FDTD) approach
\cite{Viri84,Viri86}, space and time are both discretized using
second-order finite differences. Many enhancements have been
introduced to increase the accuracy and treatment of special
boundary conditions. For example, spectral methods have been
employed to increase the accuracy of the approximation of the
spatial derivative operators \cite{Kosl82,Ezer87}; a rotated
staggered spatial grid has been developed to surmount some
instability and lack of spatial accuracy issues \cite{Saen00};
polynomial expansions of the time evolution operator have been
used to increase the accuracy and efficiency of the time
integration process \cite{Ezer87,Ezer90}.

Although each specific method cited above offers its own
advantages and drawbacks, none of these methods feature
unconditional stability and/or exact energy conservation, which
can be useful for long integration times, and high material
contrast situations, where instabilities have been noticed
\cite{Oprs99}. The algorithms that will be introduced in this
paper, address this issue.

Starting from the velocity-stress first-order differential
equations, it can be shown (see section \ref{sec:theory}) that the
solution of these equations can be written as the matrix
exponential of a skew-symmetric matrix. This constitutes an
orthogonal transformation, conserving the total energy density.
Guided by recent results regarding the numerical solution of the
time-dependent Maxwell Equations
\cite{Kole01,Kole02,HDR02_1,HDR02_2,HDR02_3,HDR02_4}, where the
underlying skew-symmetry of the equations of motion is exploited
in a matrix exponential approach, we apply this framework to the
current problem. In section \ref{sec:matrixexp} this framework is
briefly repeated for convenience and three algorithms are derived
to solve the time-dependent elastodynamic equations. The
incorporation of the presence of a source is described in section
\ref{sec:source}. For some typical examples, the performance and
efficiency of the algorithms is studied in section
\ref{sec:results}, and the conclusions are summarized in section
\ref{sec:concl}.

\section{Theory}\label{sec:theory}
In the absence of body forces, the linearized equation of momentum
conservation reads \cite{Lifs86}
\begin{equation}
\rho\frac{\partial^2}{\partial t^2}u_i=\sum_j
\frac{\partial}{\partial x_j}\sigma_{ij},
\end{equation}
where $\rho$ is the density, $u_i$ is the displacement field and
$\sigma_{ij}$ the stress field ($i=x,y,z$). It can be recast into
a coupled first order velocity-stress equation \cite{Viri84},
yielding in matrix form
\begin{equation}
\dd{t}\left(\begin{array}{c}\bsig \\
\bv\end{array}\right)=\left(\begin{array}{cc} 0 & -CD^T \\
\frac{1}{\rho}D & 0 \end{array}\right) \left(\begin{array}{c}\bsig
\\ \bv\end{array}\right). \label{mtx_nonsym}
\end{equation}
Here, $\bsig=(\sigma_{xx},\sigma_{xy},\sigma_{xz},\sigma_{yx},
\sigma_{yy},\sigma_{yz},\sigma_{zx},\sigma_{zy},\sigma_{zz})^T$),
$\bv=(v_x,v_y,v_z)^T$ is the velocity field and $D$ is the matrix
containing the spatial derivatives operators,
\begin{equation}
D=\left(\begin{array}{ccccccccc} \dd{x} & \frac{1}{2}\dd{y} &
\frac{1}{2}\dd{z} & \frac{1}{2}\dd{y} & 0 & 0 & \frac{1}{2}\dd{z}
& 0 & 0 \\ 0 & \frac{1}{2}\dd{x} & 0 & \frac{1}{2}\dd{x} & \dd{y}
& \frac{1}{2}\dd{z} & 0 & \frac{1}{2}\dd{z} & 0 \\ 0 & 0 &
\frac{1}{2}\dd{x} & 0 & 0 & \frac{1}{2}\dd{y} & \frac{1}{2}\dd{x}
& \frac{1}{2}\dd{y} & \dd{z} \end{array}\right).
\end{equation}
The stiffness tensor $C=\{C_{ijkl}\}$ relates the stress and
strain,
\begin{equation}
\sigma_{ij}=C_{ijkl}e_{kl},
\end{equation}
and is symmetric and positive definite for elastic solids.

Some important symmetries in matrix equation (\ref{mtx_nonsym})
can be made explicit by introducing the fields
\begin{equation}
\bw=\sqrt{\rho}\,\bv,
\end{equation}
and
\begin{equation}
\bs=\frac{1}{\sqrt{C}}\bsig.
\end{equation}
The expression $\sqrt{C}$ is valid since $C$ is symmetric and
positive definite.

By definition, the length of $\psi\equiv(\bs,\bw)^T$, given by
\begin{equation}
\|\psi\|^2 = \langle \psi|\psi\rangle \equiv\int_V\psi^T\psi\ d\br
= \int_V \left(\bw^2 +\bs^2 \right) d\br= \int_V \left(\rho \bv^2
+\bsig^T C^{-1}\bsig \right) d\br \label{normpsi}
\end{equation}
is related to the elastic energy density
\begin{equation}
w\equiv \frac{1}{2}\left(\rho \bv^2 +\bsig^T C^{-1}\bsig \right)
=\frac{1}{2}\left(\rho \bv^2 +
\sum_{ijkl}C_{ijkl}e_{ij}e_{kl}\right), \label{wt}
\end{equation}
of the fields.

In terms of $\psi$, matrix equation (\ref{mtx_nonsym}) becomes
\begin{equation}
\dd{t}\psi=\left(\begin{array}{cc} 0 & -
\sqrt{C}D^T\frac{1}{\sqrt{\rho}}
\\ \frac{1}{\sqrt{\rho}}D\sqrt{C} & 0
\end{array}\right)\psi\equiv \mH\psi.
\label{mtx_sym}
\end{equation}
Using the symmetric properties of $\rho$ and $\sqrt{C}$, one can
prove that the matrix $\mH$ is skew-symmetric
\begin{equation}
\mH^T = \left(\begin{array}{cc} 0 &
\left(\frac{1}{\sqrt{\rho}}D\sqrt{C}\right)^T \\
-\left(\sqrt{C}D^T\frac{1}{\sqrt{\rho}}\right)^T & 0
\end{array}\right) = \left(\begin{array}{cc} 0 &
\sqrt{C}D^T\frac{1}{\sqrt{\rho}}
\\ -\frac{1}{\sqrt{\rho}}D\sqrt{C} & 0
\end{array}\right)=-\mH,
\end{equation}
with respect to the inner product as defined in equation
(\ref{normpsi}). The formal solution of equation (\ref{mtx_sym})
is given by
\begin{equation}
\psi(t) = e^{t\mH}\psi(0)\equiv \mU(t)\psi(0), \label{eqn:formal}
\end{equation}
where $\psi(0)$ represents the initial state of the fields and the
operator $\mU$ determines their time evolution. And, since $\mH$
is skew-symmetric the time evolution operator, $\mU$ is an
orthogonal transformation:
\begin{equation}
\mU(t)^T=\mU(-t)=\mU^{-1}(t)=e^{-t\mH},
\end{equation}
and it follows that
\begin{equation}
\langle \mU(t)\psi(0)|\mU(t)\psi(0)\rangle
=\langle\psi(t)|\psi(t)\rangle= \langle\psi(0)|\psi(0)\rangle.
\end{equation}
Hence, the time evolution operator $\mU(t)$ rotates the vector
$\psi(t)$ without changing its length $\|\psi\|$. In physical
terms, this means that the total energy density of the fields does
not change with time, as can be expected on physical
grounds~\cite{Lifs86}.

\begin{figure}[t]
\begin{center}
\includegraphics{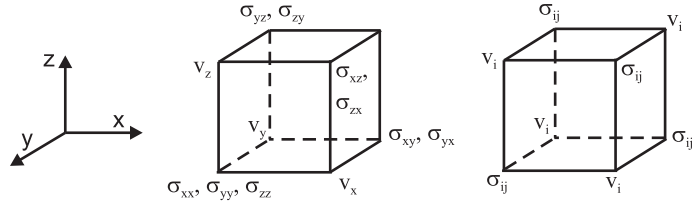}
\bigskip \caption{Unit cell of the three-dimensional staggered grid
onto which the continuous velocity and stress fields of the
elastodynamic equations are mapped in order to conserve the
skew-symmetry. Left: grid for elastic isotropic solids. Note: the
Lam\'e constants $\lambda$ and $\mu$ coincide with the stress
field components, and the mass density is only defined on velocity
field points. Right: grid for the general anisotropic case
($i,j=x,y,z$).} \label{fig1}
\end{center}
\end{figure}

In practice, the construction of a numerical algorithm requires to
discretize space and time. During both these procedures, the
skew-symmetry of $\mH$ (during the spatial discretization) and the
orthogonality of $\mU$ (during the time integration) should be
conserved. For the discretization of space, this requirement can
be met by choosing a staggered spatial grid \cite{Viri86} and a
central difference approximation for the spatial derivative. This
yields a skew-symmetric matrix $H$ for the discrete analogue of
$\mH$. A unit cell of the grid is shown in figure \ref{fig1}.
The explicit form of $H$ is derived in the appendix, in the case
of a two-dimensional isotropic elastic solid. Accordingly, the
discrete analogue of $\psi(t)$ is given by vector $\Psi(t)$.

The continuous problem, defined by $\mH$, is now translated to a
lattice problem defined by $H$:
\begin{equation}
\Psi(t)=\exp(tH)\Psi(0)\equiv U(t)\Psi(0).\label{discr_mtx_eq}
\end{equation}
or, in the time-stepping approach, we have for a small timestep
$\tau$
\begin{equation}
\Psi(t+\tau)=\exp(\tau H)\Psi(t)=U(\tau)\Psi(t).
\end{equation}
At this point, we invoke three different strategies to perform the
time integration, i.e. to approximate the matrix exponential
$\exp(tH)$. Here, we closely follow the derivation of algorithms
to solve the time-dependent Maxwell equations
\cite{Kole01,Kole02,HDR02_1,HDR02_2,HDR02_3,HDR02_4}, where the
problem to be solved is stated in a very similar form, although
the underlying physics is different. The first algorithm is based
on conserving the existing symmetries during the discretization of
time, and is unconditionally stable. Here, the time integration is
carried out by a time-stepping procedure. The second algorithm is
based on approximating the solution itself for a particular time
instance, by means of a Chebyshev expansion, and constitutes
therefore a ``one-step'' algorithm. The last algorithm is based on
recasting the original velocity-stress finite-difference algorithm
into matrix exponential form. This allows to remove the
staggered-in-time nature, and offers an elegant way to derive
higher-order in time algorithms. The construction of the
algorithms is briefly repeated in the next section.

\section{The matrix exponential approach}\label{sec:matrixexp}
\subsection{Unconditionally stable algorithms}\label{sec:uncond}

A sufficient condition for an algorithm to be unconditionally
stable is that~\cite{Smith85}
\begin{equation}
\|U(\tau)\Psi(t)\|\leq \|\Psi(t)\|. \label{UNCOND}
\end{equation}
Since $U(\tau)$ is an orthogonal transformation (see previous
section), we have $\|U(\tau)\Psi(t)\|= \|\Psi(t)\|$, and it is
sufficient to conserve the orthogonality of $U(t)$ for an
approximation $\tilde{U}(t)$ to $U(t)$, in order to construct an
unconditionally stable algorithm. One way to accomplish this is to
make use of the Lie-Trotter-Suzuki
formula~\cite{Trotter59,Suzuki77} and generalizations
thereof~\cite{Suzuki8591,DeRaedt83}. If the matrix $H$ is
decomposed, so that ${H}=\sum_{i=1}^{p}{H}_i$, then
\begin{equation}
U_1(\tau)=e^{\tau{H}_1}\ldots e^{\tau{H}_p}, \label{tsapprox}
\end{equation}
is a first order approximation to $U(\tau)$. More importantly, if
each matrix $H_i$ is skew-symmetric, then $U_1(\tau)$ is
orthogonal by construction, and hence, algorithms based on
$U_1(\tau)$, are unconditionally stable. Using the fact that both
$U(\tau)$ and $U_1(\tau)$ are orthogonal matrices, the error on
$U_1(\tau)$ is subject to the upper bound~\cite{DeRaedt87}
\begin{equation}
\|U(\tau)-U_1(\tau)\|\leq\frac{\tau^2}{2}\sum_{i<j}^p\|[{H}_i,{H}_j]\|\,,
\label{tserror}
\end{equation}
where $[{H}_i,{H}_j]=H_i H_j - H_j H_i$.

In the appendix, the decomposition of $H$ is carried out for
two-dimensional isotropic elastic solids, for which $p=12$, and it
is shown that each matrix $H_i$ is block diagonal. The computation
of the matrix exponential of a block diagonal matrix $H_i$ can be
performed efficiently, as it is equal to the block-diagonal matrix
of the matrix exponentials of the individual blocks. Therefore,
the numerical calculation of $e^{\tau H_i}$ reduces to the
calculation of matrix exponentials of $2\times2$ matrices, which
are rotations.

In practice, implementation of the first order algorithm is all
that is required to construct higher order algorithms. This is due
to the fact that in the product-formula approach, the accuracy of
an approximation can be improved in a systematic way by reusing
lower order approximations, without changing the fundamental
symmetries. For example, the orthogonal matrix
\begin{equation}
U_2(\tau)={U_1(-\tau/2)}^TU_1(\tau/2)= e^{\tau{H}_p/2}\ldots
e^{\tau{H}_1/2}e^{\tau{H}_1/2}\ldots e^{\tau{H}_p/2},
\label{secordapp}
\end{equation}
is a second-order approximation to
$U(\tau)$~\cite{Suzuki8591,DeRaedt83}. A particularly useful
fourth-order approximation (applied in for example
~\cite{Kole01,Kole02,Suzuki77,Suzuki8591,DeRaedt83,DeRaedt87,Chorin78,%
Koboyashi94,DeRaedt94,Rouhi95,Shadwick97,Krech98,%
Tran98,Michielsen98,DeRaedt00}) is given by~\cite{Suzuki8591}
\begin{equation}
U_4(\tau)=U_2(a\tau)U_2(a\tau)U_2((1-4a)\tau)U_2(a\tau)U_2(a\tau),
\label{fouordapp}
\end{equation}
where $a=1/(4-4^{1/3})$.

\subsection{One-step algorithm}\label{sec:cheby} A well-known
alternative for timestepping is to use Chebyshev polynomials to
construct approximations to time-evolution operators
\cite{Cheby,Cheby2,Cheby3,Cheby4,Cheby5}. This approach has also
been successfully applied to the problem of seismic wave
propagation \cite{Ezer87,Ezer90}. However, the main differences
between these implementations and the algorithm explained below,
besides the spatial discretization (which is based here on a
central difference approximation, instead of a spectral method),
is that in the present case, matrix $H$ is explicitly
anti-symmetrized, which gives rise to purely imaginary eigenvalues
\cite{Wilkinson}. This is an important property to justify the
validity of the expansion. In the derivation of the current
algorithm, we follow \cite{HDR02_1,HDR02_2,HDR02_3}, a recent
implementation of the Chebyshev algorithm to solve electromagnetic
wave propagation.

The basic idea is to expand the time evolution matrix
$U(t)=\exp(tH)$ for a specific time instance $t$ in matrix valued
Chebyshev polynomials on the domain of eigenvalues of $H$, which
lies entirely on the imaginary axis since $H$ is skew-symmetric.
For proper application of the expansion, the domain of eigenvalues
is rescaled to $[-1,1]$, by considering the matrix $B=-iH/\| H
\|_1$, where $\| H \|_1$ denotes the 1-norm of the matrix. It is
given by $\|H\|_1\equiv\max_j\sum_i|H_{ij}|$, see
\cite{Wilkinson}, and is easy to compute since the matrix $H$ is
sparse. Operating on state $\Psi(0)$, the expansion becomes
\begin{equation}
\Psi(t)=\exp(tH)\Psi(0)=\exp(izB)\Psi(0)= \left[J_0(z)I+
2\sum_{n=1}^{\infty} J_n(z)
\widetilde{T}_n(B)\right]\Psi(0),\label{eqn:cheb_sum}
\end{equation}
where $I$ is the identity matrix, $z=t\|H\|$, $J_n$ are $n$th
order Bessel functions and $\widetilde{T}_n(B)=i^nT_n(B)$ are the
modified Chebyshev polynomials, defined by the recursion relation
\begin{subequations}\label{cheb_recur2}
\begin{align}
\widetilde{T}_0(B)\Psi(0)&=\Psi(0), \\
\widetilde{T}_1(B)\Psi(0)&=iB\Psi(0),\\
\widetilde{T}_{n+1}(B)\Psi(0) &=
2iB\widetilde{T}_n(B)\Psi(0)+\widetilde{T}_{n-1}(B)\Psi(0)
\quad\text{for $n\geq 1$}.
\end{align}\end{subequations}
Due to the fact that the matrix $B$ is purely imaginary, it
follows from the above recursion relation (\ref{cheb_recur2}) that
$\widetilde{T}_n(B)\Psi(0)$ and thus $\Psi(t)$ will be real valued
and no complex arithmetic is involved, as should be the case.

In practice, the summation in Eq.~(\ref{eqn:cheb_sum}) will be
truncated at some expansion index $m$. This number depends on the
value of $z$, since the amplitude of the coefficients $J_n(z)$
decrease exponentially for $n>z$; this is explained in more detail
in Refs. \cite{HDR02_1,HDR02_2,HDR02_3}. Consequently, the
computation of one timestep amounts to carrying out $m$
repetitions of recursion relation Eq.~(\ref{cheb_recur2}) to
obtain the final state. This is a simple procedure: only the
multiplication of a vector with a sparse matrix and the summation
of vectors are involved.

\subsection{A modified VS-FDTD algorithm}\label{sec:mod} In section
\ref{sec:uncond}, it was shown that in the product formula
formalism, higher-order in time algorithms can be constructed by
reusing the lower order algorithms. This elegant technique to
increase the accuracy of time integration can also be applied to
FDTD algorithms (in case of the Maxwell Equations, see
\cite{HDR02_4}), and hence the conventional VS-FDTD algorithm, if
it is recast into an exponent operator form.

The update equations of the VS-FDTD \cite{Viri84,Viri86} algorithm
can be written as
\begin{equation}
\left(\begin{array}{c}\bsig(t+\tau) \\
\bv(t+\tau/2)\end{array}\right)=(I+\tau A)(I+\tau B)
\left(\begin{array}{c}\bsig(t)
\\ \bv(t-\tau/2)\end{array}\right) \equiv
U_1(\tau)\left(\begin{array}{c}\bsig(t) \\
\bv(t-\tau/2)\end{array}\right),
\end{equation}
where
\begin{equation}
A = \left(\begin{array}{cc} 0 & -CD^T \\ 0 & 0
\end{array}\right),
\end{equation}
and
\begin{equation}
B = \left(\begin{array}{cc} 0 & 0 \\ \frac{1}{\rho}D & 0
\end{array}\right).
\end{equation}
Since $A^2=B^2=0$, the time evolution operator $U_1(\tau)$ is
equal to
\begin{equation}
U_1(\tau)=\exp(\tau A)\exp(\tau B)\label{vns_1},
\end{equation}
and second-order accurate in time operating on fields that are
defined staggered-in-time. However, if $U_1(\tau)$ is interpreted
as an approximation to the operator $\exp(\tau A+\tau B)$, working
on fields non staggered-in-time, it is a first-order
approximation. Furthermore, since the time evolution operator is
now expressed in matrix exponential form, the accuracy can be
increased by the same procedure as was used for the
unconditionally stable algorithms (cf. eqs. (\ref{secordapp}) and
(\ref{fouordapp})). Therefore, the operator
\begin{equation}
U_2(\tau)=\exp(\tau A/2)\exp(\tau B)\exp(\tau A/2)\label{vns_2},
\end{equation}
constitutes a second-order approximation to the exact time
evolution operator. And similarly, a fourth order algorithm can be
derived, see Eq.~(\ref{fouordapp}).

So, by introducing a small modification to the original VS-FDTD
algorithm, that would require a minimal change in existing
numerical codes, the staggered-in-time nature is removed, and
higher-order in time algorithms are derived.

\section{Sources}\label{sec:source}
In the presence of an explosive initial condition, or other
time-dependent body force, the equation of motion reads
\begin{equation}
\dd{t}\psi(t)=\mH\psi(t)+\phi(t),
\end{equation}
where the time-dependent source is denoted by the term $\phi(t)$.
The formal solution is given by
\begin{equation}
\psi(t)=\exp(t\mH)\psi(0)+\int_0^{t}\exp((t-u)\mH)\phi(u)du.\label{sol_source}
\end{equation}
In the time-stepping approach (e.g. the unconditionally stable
algorithms), the source term --if its time dependence is known
explicitly-- can be integrated for each timestep. For example, a
standard quadrature formula can be employed to compute the
integral over $u$, like the fourth-order accurate Simpson rule
\cite{Atkinson}
\begin{equation}
\int_t^{t+\tau} e^{(t+\tau-u)\mH}\phi(u)\,du \approx
\frac{\tau}{6}\left( e^{\tau H}\phi(t)+4e^{\tau
H/2}\phi(t+\tau/2)+\phi(t+\tau)\right).\label{simpson}
\end{equation}

In case of the one-step algorithm, this approach of incorporating
the source term is not efficient, as for each value of $t-u$ the
recursion (\ref{cheb_recur2}) would have to be performed, and a
different route is taken. Instead, each of the two terms on the
right hand side of equation (\ref{sol_source}) is expanded in
modified Chebyshev polynomials separately. The expansion of the
first term is discussed in section \ref{sec:cheby}. For the second
term, the source-term integral, the procedure is carried out here
for a source with Gaussian time dependence,
\begin{equation}
\phi(t)=f(t)S(\br)=\exp(-\alpha(t-t_0)^2)S(\br),\label{src:gaussian}
\end{equation}
where $S(\br)$ denotes the spatial dependence of the source.

The source term
\begin{equation}
h(t,t_0,\alpha,H)=\int_0^t du e^{(t-u)H}f(u),
\end{equation}
is expanded in modified Chebyshev polynomials,
\begin{equation}
h(t,t_0,\alpha,H)S(\br)=\left(\frac{1}{2}b_0I+\sum_{k=0}^K b_k
\widetilde{T}_k\right)S(\br),
\end{equation}
where the expansion coefficients are given by
\begin{equation}
b_k=i^{-k} \frac{2}{\pi}\int_0^{\pi}h(t,t_0,\alpha,\cos\theta)\cos
k\theta d\theta. \label{cheb_coeff_source}
\end{equation}
The replacement of $H$ by $\cos\theta$ emphasizes that $H$ should
be normalized such that all eigenvalues lie in the range $[-1,1]$.
We proceed by evaluating $h(t,t_0,\alpha,H)$. After substitution
of $z=\|H\|_1$, $z_0=t_0\|H\|_1$, $\beta=\alpha/\|H\|_1^2$ and
$x=-iH/\|H\|_1$ normalize the matrix $H$, we obtain
\begin{equation}
h(z,z_0,\beta,x) = \frac{1}{2}\sqrt{\frac{\pi}{\alpha}}
\exp\left((z-z_0)ix-x^2/4\beta\right)
\left[\erf\left(z_0\sqrt{\beta}-\frac{ix}{2\sqrt{\beta}}\right)+
\erf\left((z-z_0)\sqrt{\beta}+\frac{ix}{2\sqrt{\beta}}\right)\right].
\end{equation}
Now we put $x=\cos\theta$ and the remaining integral over $\theta$
in equation (\ref{cheb_coeff_source}) is computed by a Fast
Fourier transformation:
\begin{equation}
b_k=2i^{-k}\sum_{n=0}^{N-1}e^{2\pi ink/N}h(z,z_0,\beta,\cos
\frac{2\pi n}{N}).
\end{equation}

The derivation of the Chebyshev expansion coefficients for a
source defined by equation
\begin{equation}
g(t)=\dd{t}f(t)=-2\alpha
(t-t_0)\exp(-\alpha(t-t_0)^2).\label{src:dgaussian}
\end{equation}
is very similar, and will not be treated explicitly here.

\section{Results}\label{sec:results}
The performance and accuracy of the algorithms introduced in the
previous sections is studied by comparing the results with a
reference solution generated by the one-step algorithm, denoted by
$\hat{\Psi}(t)$. This choice is motivated by the fact that the
latter, considering the time integration, produces numerically
exact results \cite{Ezer87,Ezer90,Cheby2}. Furthermore, there are
rigorous bounds on the error of the unconditionally stable
algorithm (cf. Eq.~(\ref{tserror})): in the presence of a source
$\phi(t)$, the difference between the exact solution $\Psi(t)$ and
the approximate solution $\tilde{\Psi}(t)$, obtained by using the
4th order unconditionally stable algorithm, is bounded by
\cite{HDR02_4}
\begin{equation}
\| \Psi(t)-\tilde{\Psi}(t)\| \leq c_4 t\tau^4 \left(\| \Psi(0)
\|+\int_0^t du \| \phi(u) \|\right),\label{ut4bound}
\end{equation}
where $c_4$ is a constant. For the difference between the exact
solution and the solution obtained by using the one-step
algorithm, we can write using the triangle inequality
\begin{equation}
\| \Psi(t)-\hat{\Psi}(t)\| \leq \| \Psi(t)-\tilde{\Psi}(t)\| + \|
\tilde{\Psi}(t)-\hat{\Psi}(t)\|.\label{triang}
\end{equation}
Using equations (\ref{ut4bound}) and (\ref{triang}) and the fact
that the difference $\| \tilde{\Psi}(t)-\hat{\Psi}(t) \|$ vanishes
with $\tau^4$, as we will show below, we can be confident that the
one-step algorithm indeed produces numerically exact results. This
justifies to define the error as $\|
\tilde{\Psi}(t)-\hat{\Psi}(t)\|/\| \hat{\Psi}(t)\|)$.

The performance of the following algorithms is considered: the
original VS-FDTD algorithm, denoted by Vir; the unconditionally
stable algorithms, denoted by LTS-2 and LTS-4, for resp. second
and fourth order accuracy in time; the non staggered-in-time,
modified VS-FDTD algorithm, denoted by VNS-2 and VNS-4, depending
on the accuracy in time.

\begin{figure}[t]
\begin{center}
\includegraphics{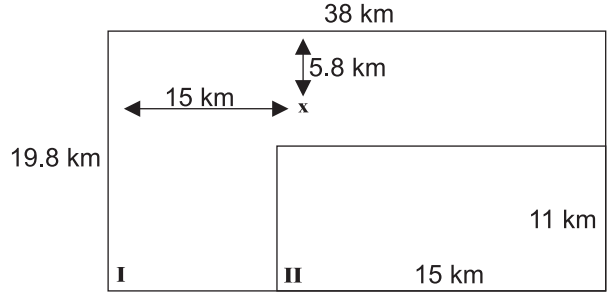}
\bigskip\caption{The corner-edge system, consisting of two
different materials. The system size and location of the source
({\bf x}) are indicated in the picture. At the top, a free-surface
boundary condition is imposed, the other boundaries are rigid. The
overall density is $\rho=2500$ kgm$^{-3}$, and the mesh size is
$\delta=100$ m. In the bulk material (I), the wave velocities are
$v_p=6$ kms$^{-1}$ and $v_s=2$ kms$^{-1}$, whereas in the inner
material (II), they are $v_p=9$ kms$^{-1}$ and $v_s=3$ kms$^{-1}$.
The source excites the $s_{xx}$ and $s_{yy}$ stress fields with
time dependence $g(t)$ from equation (\ref{src:dgaussian}) and
parameters $\alpha=40$ and $t_0=1.5$ s.}
\label{fig:corner_edge_setup}
\end{center}
\end{figure}

\begin{figure}[t]
\begin{center}
\includegraphics{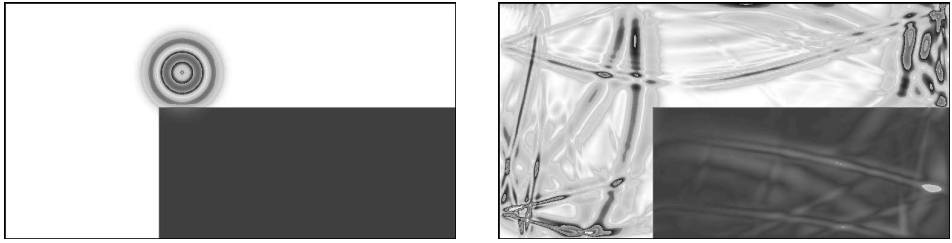}
\bigskip\caption{Two snapshots of the kinetic-energy density distribution
of the corner-edge system of figure \ref{fig:corner_edge_setup}.
Left: state at $t=1.8$ s, right: state at $t=6.0$ s.}
\label{fig:corner_edge_sim}
\end{center}
\end{figure}

Consider a rectangular system consisting of two different
materials, displayed in figure \ref{fig:corner_edge_setup}.
This system is also studied in references \cite{Viri86,Kell76},
and proved to be a good testing situation for the performance of
an algorithm solving the elastodynamic equations. The time
evolution of the velocity and stress fields is computed using an
explosion as initial condition, modeled by equation
(\ref{src:dgaussian}), up to $t=6$ s, with the six different
algorithms. In figure \ref{fig:corner_edge_sim}, the
kinetic-energy density distribution is shown for two different
time instances.

\begin{figure}[t]
\begin{center}
\includegraphics{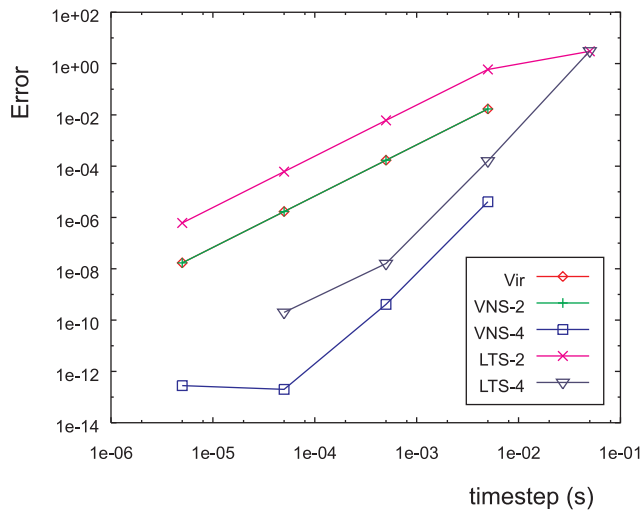}
\bigskip\caption{%
Error as function of timestep for all timestepping algorithms, for
the system defined in figure \ref{fig:corner_edge_setup}, using
the data from table \ref{tbl:corner_edge_results}.}
\label{fig:corner_edge_results}
\end{center}
\end{figure}

For the corner-edge system, the errors are listed in table
\ref{tbl:corner_edge_results}, and are shown in figure
\ref{fig:corner_edge_results}.
\begin{table}[t]
\begin{center}
\begin{tabular}{|c|c|c|c|c|c|}
Time-step (s) & Vir & VNS-2 & VNS-4 & LTS-2 & LTS-4 \\ \hline
$5\ee{-2}$ & $\infty$     & $\infty$     & $\infty$      & $3.0\ee{0}$  & $3.1\ee{0}$ \\
$5\ee{-3}$ & $1.7\ee{-2}$ & $1.7\ee{-2}$ & $4.1\ee{-6}$  & $5.9\ee{-1}$ & $1.6\ee{-4}$ \\
$5\ee{-4}$ & $1.7\ee{-4}$ & $1.7\ee{-4}$ & $4.1\ee{-10}$ & $6.1\ee{-3}$ & $1.6\ee{-8}$ \\
$5\ee{-5}$ & $1.7\ee{-6}$ & $1.7\ee{-6}$ & $2.0\ee{-13}$ & $6.1\ee{-5}$ & $2.0\ee{-10}$ \\
$5\ee{-6}$ & $1.7\ee{-8}$ & $1.7\ee{-8}$ & $2.8\ee{-13}$ & $6.1\ee{-7}$ & - \\
\end{tabular}
\bigskip
\caption{%
Error as function of timestep for all timestepping algorithms, for
the system defined in figure \ref{fig:corner_edge_setup}. An
infinite symbol ($\infty$) denotes that the algorithm was not
stable, whereas in one case (-) the computation was not performed.
The error in the staggered-in-time Vir algorithm is determined by
averaging the error in the kinetic and potential energy density at
respectively the final time instance and the final time instance
shifted by half a timestep. The time shifting procedure is carried
out by the Chebyshev algorithm and also applied to prepare the
initial condition.} \label{tbl:corner_edge_results}
\end{center}
\end{table}
For the largest timestep, the VS-FDTD algorithms (Vir,VNS-2,VNS-4)
are unstable. This can be expected, since the maximum timestep is
limited by the largest velocity, the $v_p$ velocity, and the
mesh-size, through the Courant limit \cite{Viri86}
\begin{equation}
\tau < \frac{\delta}{v_p\sqrt{2}}.
\end{equation}
Furthermore, from table \ref{tbl:corner_edge_results} it is clear
that for all algorithms the error scales according to the order of
accuracy in time. We also see that for the corner-edge system, the
VS-FDTD algorithms perform much better than the energy-conserving
LTS algorithms, as long as the timestep is smaller than the
Courant limit. For timesteps larger than the Courant limit, the
LTS algorithms (LTS-2 and LTS-4) are stable, although the error
does not (yet) scale according to the order of accuracy in time
\cite{higher_order_ref}. For very accurate results (errors below
$10^{-10}$), the number of operations the achieve this accuracy
becomes so large that the error does not scale systematic anymore.

With respect to the efficiency of the algorithms, we note that the
number of matrix-vector operations $W$, necessary to perform one
timestep, is 1 for the Vir algorithm, 1.5 for the second-order
VNS-2 and LTS-2 algorithms, and 10 for the fourth-order VNS-4 and
LTS-4 algorithms. For the specific example here, the corner-edge
system, the one-step algorithm employs $m=1514$ expansion terms.
At $t=6$ and $\tau=0.005$, the VNS-2 algorithm already uses more
(namely 1800) matrix-vector operations. Therefore, we draw the
conclusion that one-step algorithm should be preferred to be used
to solve the time evolution for this problem. Note that in
general, the choice of which algorithm to use depends heavily on
which degree of error is acceptable. In this specific case, there
are values for $\tau$ for which the VNS-2 algorithm uses less
matrix-vector operations than the one-step algorithm, but then the
error will be larger than the error for $\tau=0.005$, and maybe
unacceptably high. On the other hand, one VNS-2 or LTS-2
matrix-vector operation is carried out (in practice) faster than
one Chebyshev recursion iteration, although this depends on the
actual implementation.

\begin{figure}[t]
\begin{center}
\includegraphics{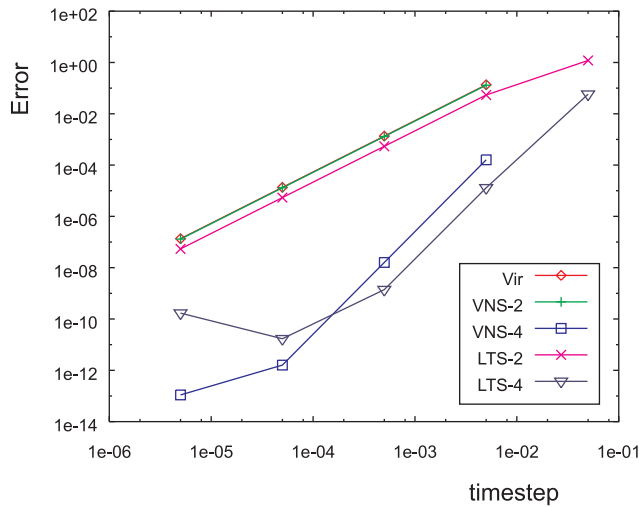}
\bigskip\caption{%
Error as function of timestep for all timestepping algorithms, for
a random medium and random initial conditions, using the data from
table \ref{tbl:seismic_random}.} \label{fig:seismic_random}
\end{center}
\end{figure}

It is important to note that the initial condition plays an
important role in the error of the solution produced by a specific
algorithm. From the results of the corner-edge system, one might
draw the conclusion that the VS-FDTD algorithms achieve better
results than the LTS algorithms for all systems. This is not true.
In table \ref{tbl:seismic_random}, the error is listed as function
of timestep for all algorithms, as compared with the one-step
algorithm, for a system consisting of a random medium (also
studied in for example \cite{Kneip93}) and starting from a random
initial condition. The results are also shown in figure
\ref{fig:seismic_random}. From the table and the figure, it is
clear that in this case, the LTS family of algorithms perform
better than the VS-FDTD algorithms. Again we see that the error
scales according to the order of accuracy in time, except for very
small errors (below $10^{-10}$). Especially for the LTS-4
algorithm, we see that accumulation of rounding errors, due to
large number of operations, increases the error.
\begin{table}[t]
\begin{center}
\begin{tabular}{|c|c|c|c|c|c|}
Time-step (s) & Vir & VNS-2 & VNS-4 & LTS-2 & LTS-4 \\ \hline
$5\ee{-2}$ & $\infty$     & $\infty$     & $\infty$      & $1.2\ee{0}$ & $5.8\ee{-2}$ \\
$5\ee{-3}$ & $1.3\ee{-1}$ & $1.3\ee{-1}$ & $1.6\ee{-5}$  & $5.4\ee{-2}$ & $1.3\ee{-5}$ \\
$5\ee{-4}$ & $1.3\ee{-3}$ & $1.3\ee{-3}$ & $1.6\ee{-9}$  & $5.4\ee{-4}$ & $1.4\ee{-9}$ \\
$5\ee{-5}$ & $1.3\ee{-5}$ & $1.3\ee{-5}$ & $1.6\ee{-13}$ & $5.4\ee{-6}$ & $1.7\ee{-11}$ \\
$5\ee{-6}$ & $1.3\ee{-7}$ & $1.3\ee{-7}$ & $1.1\ee{-13}$ & $5.4\ee{-8}$ & $1.7\ee{-10}$ \\
\end{tabular}
\bigskip \caption{%
Error as function of timestep for all timestepping algorithms. In
this case, the one-step algorithm employs $m=321$ expansion terms.
The system measures $L_x=L_y=10$, with a mesh of $\delta=0.1$, and
the material parameters $\rho,\lambda,\mu$ vary randomly in space
with values distributed randomly in the interval $[1,3]$. The
error is determined at $t=3$. (all quantities are expressed here
in dimensionless units).} \label{tbl:seismic_random}
\end{center}
\end{table}

\section{Conclusions}\label{sec:concl}
In this paper, we have introduced algorithms to solve the
time-dependent elastodynamic equations, based on a matrix
exponential approach. The conservation of the underlying
skew-symmetry of the first-order partial differential equations
while discretizing the spatial operators and fields, offers a
sound starting point to expand the time evolution operator in
Chebyshev polynomials. The resulting one-step algorithm is
accurate up to machine precision, and this statement is justified
by rigorous bounds on the error of the unconditionally stable
algorithms. The latter class of algorithms proved particularly
useful if the total energy should be conserved or if a random
initial condition is used.

Finally, the original VS-FDTD algorithm is modified by recasting
it into an exponent operator form. In this new formulation, the
staggered-in-time nature is removed and higher-order in time
algorithms are derived, based on the lower-order algorithms.
Existing VS-FDTD codes can be modified with minor effort to
benefit from these advantages.

In this paper, only free and rigid boundary conditions are
considered. Future research is aimed at incorporating absorbing
boundary conditions, and the presence of visco-elastic materials.
More sophisticated discretization schemes, conserving the
skew-symmetry, can be easily incorporated \cite{Kole02}, and do
not require conceptual changes.

\section*{Acknowledgements}
The author acknowledges H. De Raedt, M.T. Figge and K. Michielsen
for useful discussions. This work is partially supported by the
Dutch `Stichting Nationale Computer Faciliteiten' (NCF).

\section*{Appendix: discretization and decomposition of $\mH$ for
two-dimensional isotropic elastic solids}\label{app_discrH}

\begin{figure}[t!]
\begin{center}
\includegraphics{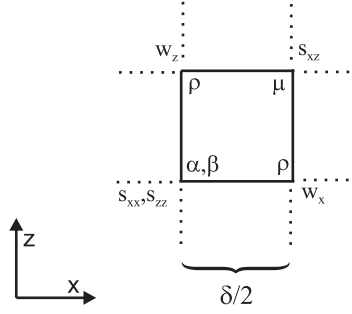}
\bigskip\caption{Unit cell of the two dimensional staggered grid onto
which the continuous velocity and stress fields of the
elastodynamic equations are mapped in order to conserve the
skew-symmetry.} \label{2d_grid}
\end{center}
\end{figure}

In this appendix, it is shown how the matrix
\begin{equation}
\mH=\left(\begin{array}{cc} 0 & -
\sqrt{C}D^T\frac{1}{\sqrt{\rho}}
\\ \frac{1}{\sqrt{\rho}}D\sqrt{C} & 0
\end{array}\right),
\end{equation}
is discretrized conserving its skew-symmetric properties, for two
dimensional isotropic elastic solids. In two dimensional P-SV wave
propagation, no dependency upon $y$ is assumed. For isotropic
elastic solids, the stiffness matrix is given in terms of the
Lam\'e coefficients $\lambda$ and $\mu$ by \cite{Lifs86}
\begin{equation}
C=\left(\begin{array}{ccc} \lambda+2\mu & \lambda & 0 \\ \lambda &
\lambda+2\mu & 0 \\ 0 & 0 & \mu \end{array}\right),
\end{equation}
in the basis $\bsig=(\sigma_{xx},\sigma_{zz},\sigma_{xz})^T$. In
the skew-symmetric basis, using the variables
$\bw=\sqrt{\rho}\,\bv$ and $\bs=C^{-1/2}\bsig$, one needs
$\sqrt{C}$, which reads
\begin{equation}
\sqrt{C}=\left(\begin{array}{ccc} \alpha & \beta & 0 \\ \beta &
\alpha & 0 \\ 0 & 0 & \sqrt{\mu} \end{array}\right),
\end{equation}
where
\begin{align}
\alpha &=
\frac{1}{\sqrt{2}}\left(\sqrt{\lambda+\mu}+\sqrt{\mu}\right), \\
\beta  &=
\frac{1}{\sqrt{2}}\left(\sqrt{\lambda+\mu}-\sqrt{\mu}\right).
\end{align}
This gives for the explicit form of the matrix $\mH$
\begin{equation}
\mH=\left(\begin{array}{ccc|cc}
0 & 0 & 0 & \alpha\dd{x}\fsr & \beta\dd{z}\fsr \\
0 & 0 & 0 & \beta\dd{x}\fsr & \alpha\dd{z}\fsr \\
0 & 0 & 0 & \sqrt{\mu}\dd{z}\fsr & \sqrt{\mu}\dd{x}\fsr \\
\hline
\fsr\dd{x}\alpha & \fsr\dd{x}\beta & \fsr\dd{z}\sqrt{\mu} & 0 & 0 \\
\fsr\dd{z}\beta & \fsr\dd{z}\alpha & \fsr\dd{x}\sqrt{\mu} & 0 & 0 \\
\end{array}\right),\label{mtx_explicit}
\end{equation}
in the basis $\psi=(s_{xx},s_{zz},s_{xz},w_x,w_z)^T$. The discrete
analogue of $\psi$, the vector $\Psi$, is obtained by mapping the
fields onto the two-dimensional staggered grid \cite{Viri86} that
is shown in figure \ref{2d_grid}.
We adopt the convention that the $s_{xx}$ and $s_{zz}$ stress
fields are located in the $(1,1)$ corner of the grid. The values
of the discretized fields $f$ are related to their continuous
counterparts $g$ by
\begin{equation}
f(i,j,t)=g(i\delta/2,j\delta/2,t).
\end{equation}
Therefore, using the unit vector $\be(i,j,k)$, fields within the
vector $\Psi$ can be indexed on the grid by
\begin{equation}
\Psi(i,j,s_{xx},t)\equiv\be^T_{i,j,s_{xx}}\Psi(t)\equiv
s_{xx}(i,j,t),
\end{equation}
and analogous equations apply for indexing the other stress and
velocity fields within $\Psi$.

Due to the staggered nature of the grid and the choice of the
origin, the $s_{xx}$ and $s_{zz}$ stress fields are only defined
on the $x=$odd and $z=$odd lattice points. Similarly, the $w_x$
and $w_z$ velocity fields are defined at respectively the
$x=$even/$z=$odd and $x=$odd/$z=$even lattice points, and the
$s_{xz}$ stress field is given at the $x=$even and $z=$even grid
entries. Note that all for simplicity of notation the fields are
indexed on the full grid, despite the fact that they are not
defined on each point.

It is assumed that the total number of lattice points in each
direction is odd, and also that the boundary is located at the
first and last rows/columns of the grid. The free or rigid
boundary conditions themselves are implemented by excluding the
field points that are located at the boundary and should remain
zero during the time integration.

Using this grid and the central-difference approximation to the
spatial derivative, we obtain the spatially discretized analogue
of equation (\ref{mtx_explicit}), for example
\begin{equation}
\dd{t}s_{xx}(i,j,t)=
\alpha(i,j)\frac{1}{\delta}\left[\frac{w_x(i+1,j,t)}{\sqrt{\rho(i+1,j)}}-
\frac{w_x(i-1,j,t)}{\sqrt{\rho(i-1,j)}} \right]
+\beta(i,j)\frac{1}{\delta}\left[
\frac{w_z(i,j+1,t)}{\sqrt{\rho(i,j+1)}}-\frac{w_z(i,j-1,t)}{\sqrt{\rho(i,j-1)}}\right],
\end{equation}
and
\begin{align}
\dd{t}w_{x}(i,j,t) &= \frac{\alpha(i+1,j)s_{xx}(i+1,j,t)-
\alpha(i-1,j)s_{xx}(i-1,j,t)}{\delta\sqrt{\rho(i,j)}} \nonumber\\
&+ \frac{\beta(i+1,j)s_{zz}(i+1,j,t)-
\beta(i-1,j)s_{zz}(i-1,j,t)}{\delta\sqrt{\rho(i,j)}} \nonumber \\
&+ \frac{\sqrt{\mu(i,j+1)}
s_{xz}(i,j+1,t)-\sqrt{\mu(i,j-1)}(s_{xz}(i,j-1,t)}{\delta\sqrt{\rho(i,j)}}.
\end{align}
Similar equations hold for $s_{zz},s_{xz}$ and $w_z$. Using this
notation, the matrix $H$ can be decomposed into
\begin{equation}
H = H^{(x,s_{xx},w_x)}+H^{(x,s_{zz},w_x)}+H^{(x,s_{xz},w_z)}+
H^{(z,s_{xx},w_z)}+H^{(z,s_{zz},w_z)}+ H^{(z,s_{xz},w_x)},
\label{eqn:sepH}
\end{equation}
where, for instance, the explicit form of $H^{(x,s_{xx},w_x)}$ is
given by
\begin{align}
H^{(x,s_{xx},w_x)}&=\sump_{j=1}^{n_y}\sump_{i=1}^{n_x-2}
\frac{\alpha(i+1,j)}{\delta\sqrt{\rho(i,j)}}\left[
\be^{\pht}_{i,j,s_{xx}}\be^T_{i+1,j,w_x}-\be^{\pht}_{i+1,j,w_x}\be^T_{i,j,s_{xx}}\right]
\nonumber \\ &+ \sump_{j=1}^{n_y}\sump_{i=2}^{n_x-1}
\frac{\alpha(i,j)}{\delta\sqrt{\rho(i+1,j)}}\left[
\be^{\pht}_{i,j,s_{xx}}\be^T_{i+1,j,w_x}-\be^{\pht}_{i+1,j,w_x}\be^T_{i,j,s_{xx}}\right]
\\ &= H_1^{(x,s_{xx},w_x)} +
H_2^{(x,s_{xx},w_x)}.\label{eqn:seismic_matr_sep}
\end{align}
Here, the prime in the summation indicates that the summation
index is increased with strides of two. It is easy to convince
oneself that the matrices $H_1^{(x,s_{xx},w_x)}$ and
$H_2^{(x,s_{xx},w_x)}$ are block diagonal and skew-symmetric. The
other matrices in equation (\ref{eqn:sepH}) have similar explicit
forms and can also be decomposed into block diagonal parts.
Therefore, a first-order approximation to the matrix exponent, as
given by equation (\ref{tsapprox}), reads
\begin{gather}
\exp(\tau H) = \nonumber\\ \quad\quad \phantom{\cdot}
\exp\left(\tau H_1^{(x,s_{xx},w_x)}\right)
\exp\left(\tau H_2^{(x,s_{xx},w_x)}\right)
\exp\left(\tau H_1^{(x,s_{zz},w_x)}\right)
\exp\left(\tau H_2^{(x,s_{zz},w_x)}\right) \nonumber\\ \quad\quad \cdot
\exp\left(\tau H_1^{(x,s_{xz},w_z)}\right)
\exp\left(\tau H_2^{(x,s_{xz},w_z)}\right)
\exp\left(\tau H_1^{(z,s_{xx},w_z)}\right)
\exp\left(\tau H_2^{(z,s_{xx},w_z)}\right) \nonumber\\ \quad\quad \cdot
\exp\left(\tau H_1^{(z,s_{zz},w_z)}\right)
\exp\left(\tau H_2^{(z,s_{zz},w_z)}\right)
\exp\left(\tau H_1^{(z,s_{xz},w_x)}\right)
\exp\left(\tau H_2^{(z,s_{xz},w_x)}\right)
\nonumber\\ \quad\quad
+ \mO(\tau^2).
\label{eqn:seismic_LTS}
\end{gather}


\begin{thebibliography}{99}
\bibitem{higher_order_ref}
This behavior is also noticed when unconditionally stable
algorithms are used to solve the time-dependent Maxwell Equations,
see \cite{Kole02}.
\end{thebibliography}
\end{document}